\documentclass{llncs}
\usepackage[german,english]{babel}
\usepackage{times,theorem}
\usepackage{latexsym,color,graphics}
\usepackage{url}
\usepackage{epsfig}
\usepackage{amssymb}
\usepackage{amsmath}
\usepackage{amsfonts}

\begingroup
\catcode`\~=11
\gdef\urltilde{\lower 0.6ex\hbox{~}}
\endgroup

\newcommand{\A}{\mathcal{A}} 
 \newcommand{\D}{\mathcal{D}}
 
\newcommand{\G}{\mathcal{G}} 
 
 \renewcommand{\L}{\mathcal{L}}
\newcommand{\M}{\mathcal{M}} \newcommand{\N}{\mathcal{N}}
 \renewcommand{\P}{\mathcal{P}}
 \newcommand{\R}{\mathcal{R}}
 \newcommand{\T}{\mathcal{T}}
 \newcommand{\V}{\mathcal{V}}
\newcommand{\W}{\mathcal{W}}

\title{Binary Sequent Calculi for Truth-invariance Entailment of Finite Many-valued Logics}
\author{Zoran Majki\'c}
\institute{International Society for Research in Science and Technology \\
PO Box 2464 Tallahassee, FL 32316 - 2464 USA\\
\email{majk.1234@yahoo.com}\\ http://zoranmajkic.webs.com/}
\authorrunning{Zoran Majki\'c}

\newtheorem{theo}{Theorem}
\newtheorem{propo}{Proposition}
\newtheorem{coro}{Corollary}

\newcount\pdfoutput
\begin{document}


\maketitle

\begin{abstract}
In this paper we consider the class of truth-functional many-valued
logics with a  finite set of truth-values. The  main result of this
paper is the development of a new \emph{binary} sequent calculi
(each sequent is a pair of formulae) for  many valued logic with a
finite set of  truth values, and of  Kripke-like semantics for it
that is both sound and complete. We did not use the logic entailment
based on matrix with a strict subset of designated truth values, but
a different new kind of semantics based on the generalization of the
classic 2-valued truth-invariance entailment. In order to define
this non-matrix based sequent calculi, we transform many-valued
logic into positive 2-valued multi-modal logic with classic
conjunction, disjunction and finite set of  modal connectives. In
this algebraic framework we define an uniquely determined axiom
system, by extending the classic 2-valued distributive lattice logic
(DLL) by a new set of sequent axioms for many-valued logic
connectives. Dually, in an \emph{autoreferential} Kripke-style
framework we obtain a uniquely determined frame, where each possible
world is an equivalence class of  Lindenbaum algebra for a
many-valued logic as well, represented by a truth value.
\end{abstract}

\section{Introduction}
 A significant number of real-world applications in Artificial
Intelligence has
 to deal with partial, imprecise and uncertain information, and that
 is the principal reason for introducing  non-classic truth-functional many-valued
 logics, as for example, the fuzzy,  bilattice-based and
 paraconsistent logics, etc..
   The first formal semantics for modal logic was based on many-valuedness, proposed by Lukasiewicz
  in 1918-1920, and consolidated in 1953 in a 4-valued system of modal logic \cite{Luka53}.
  All cases of  many-valued logics mentioned above are based on
  a lattice $(X,\leq)$ of  truth values.\\
  Sequent calculus, introduced by Gentzen \cite{Gent32} and Hertz
\cite{Hert29} for classical logic, was generalized to the
many-valued case by Rouseau \cite{Rous67} and others. The tableaux
calculi were presented in \cite{Carni87,Hanh91}.
   The standard two-sides sequent
  calculi for  lattice-based many-valued logics
    has been  elaborated  recently (with an autoreferential Kripke-style semantics for such
  logics) in two complementary ways in
  \cite{Majk06ml,Majk08dC,Majk11,Majk06th} as well.\\
  In what follows we will consider,   for
a given  logic language, with a set of ground formulae (without
variables) $\L$, with a set of many-valued logic connectives in
$\Sigma$ and predicate (or propositional) letters $p,q,r,..$, the
truth-functional algebraic semantics.
 A many-valued valuation
$v:\L\rightarrow X$ in this case is a homomorphism between the free
syntax-algebra of this logic language and the algebra $(X, \Sigma)$
of truth-values. That is, for any n-ary logic connective $\odot$ in
$\Sigma$, $~\odot:X^n \rightarrow X$, it holds that
$v(\odot(\phi_1,...,\phi_n) = \odot(v(\phi_1),...,v(\phi_n))$, where
$\phi_i \in \L$, $1 \leq i \leq n$ are logic formulae in $\L$.
Notice that there is a non truth-functional many-valued semantic
presented in \cite{Avro03} as well. It is based on
"nondeterministic" connectives $\widetilde{\odot}:X^n \rightarrow
\textbf{2}^X~$ (where $A^B$ denotes the set of functions from $B$ to
$A$, and $\textbf{2} = \{0,1\}$ is a complete lattice of classical
2-valued logic) for each ordinary connective $\odot \in \Sigma$,
such that $v(\odot(\phi_1,...,\phi_n) \in
\widetilde{\odot}(v(\phi_1),...,v(\phi_n))$, and so called,
Nmatrices (non-deterministic matrices). But in this case the
compositional property (the homomorphic property above) of a
many-valued non-deterministic valuation $~v~$ is not valid. Thus, in
what follows we will consider only the truth-functional
many-valued logics where the valuations are homomorphic.\\
The well-known
semantics of the many-valued logics is based on algebraic matrices
$(X,D)$ where $D \subset X = \{x_1,...,x_m\}$ is a strict subset of
designated truth values, so that a valuation $v$ is a model for a
formula $\phi \in \L$ iff $v(\phi) \in D$. This semantics is
commonly used in practice, particularly when the number of
truth-values is limited. As in the case of three-valued
propositional logic, two different choices of the set of designated
values for the same semantics give respectively Kleene logic, and a
basic paraconsistent logic $J_3$ -both quite important for
mathematical logics itself and for its applications. \\
%
   The well known sequent system developed for such a many-valued
   logic with matrix-based semantics
is an ad hoc system based on m-sequents.
based on standard matrices.  The more detailed information for
interested readers can be found in \cite{BFZa93,BFSa00}.
But such an approach, through  absolutely correct and useful,  has
some little drawbacks, that hopefully may be improved:
\begin{itemize}
  \item It is much less known than those which are based on ordinary
two-sided sequents. The framework for two-sided sequent calculi is
well-understood, and a lot of programs have been made towards
developing their efficient implementations.
  \item The use of two-sided sequents reflects the basic fact that  logic
is all about consequence relations, while in the m-sequent calculus
only some characterization of the consequence relation can be done
in a roundabout way.
  \item The use of two-sided sequents is universal and independent of
any particular semantics, while the use of m-sequents relies in a
particular way of a specific semantics for a given logic.
\end{itemize}
 Consequently, it is interesting to consider a
 calculus for  many-valued logics based on standard
\emph{binary sequents}.
 The previous work in this direction  is recently proposed in \cite{ABNK06} based on m-valued Nmatrices,
signed formulae and a Rasiowa-Sikorski deduction system
\cite{RaSi70,Koni02}. In the approach used in \cite{ABNK06}, an
m-sequent calculus is transformed into the ordinary   two-sided
sequent system (where each sequent is of the form $\Gamma \vdash
\triangle$, where $\Gamma,
\triangle \subset \L$ are the finite subsets of logic formulae).\\
Notice that both sequent systems above are based on \emph{matrix
semantics} for logic entailment. This is not our case: we will use
the \emph{truth-invariance} semantics for many-valued logic
\cite{MajkC04} that is different
from matrix-based semantics. \\
In this paper we propose this new approach to the semantics of
many-valued logics by transforming the original many-valued logic
into the 2-valued multi-modal logic \cite{Majk06MV}, and then, based
on the classical 2-valued distributive lattice logic (DLL)
\cite{Dunn95} extended by a set of new axioms for this 2-valued
transformation of many-valued logic, by applying Dunn's
binary-sequent approach. In the original Dunn,s approach each
sequent is of the simple form $\Phi \vdash \Psi$ (here $\Phi$ and
$\Psi$ are the 2-valued multi-modal formulae) and, consequently, his
system is a particular sequent calculus where the left side of a
sequent is not generally \emph{a set} of 2-valued formulae but a
single formula. Another differences between previous approaches and
the approach used in this work are as follows:
\begin{itemize}
  \item We will not use the many-valued Rosser-Turquette
  operators $J_k$  \cite{RoTu52} that have been introduced a lot of time before
 the appearance of the
Kripke semantics for algebraic modal operators, and we will replace
them by modal operators as follows: here we will adopt the
ontological encapsulation of many-valued logic into 2-valued
multi-modal logic \cite{Majk06MV} with algebraic
modal operators $[x]:X \rightarrow \textbf{2}$ for any $x \in X$. In
what follows we will use $[x]$ both as modal operator, i.e., a
syntactic language entity for modality "heaving a truth-value $x$"
(it express exactly the \emph{modality} of truth of the formula
$\phi$), and as a function on truth values, i.e., a semantic entity,
which will be clear from the particular context where $[x]$ is used.
\item We avoided to use the signed
 formulae used in \cite{ABNK06}. In this way,
  by using modal approach and its standard Kripke semantics, we are not obligated to develop an unnecessary
 ad-hoc semantics for such a calculus as it will be shown in Section \ref{Section:Kripke}.
\end{itemize}
 Notice that the main result of this work is  the fact that we  obtained a standard \emph{binary} sequent calculi for finite
 many-valued logic with the \emph{truth-invariance  semantics}
 of logic
 entailment. Consequently, this work is not only another new reformulation of the same many-valued inference system based on matrices,  but is
 substantially  new inference system, different from the m-sequents and from the sequent calculi
 in \cite{ABNK06} (that are mutually equivalent).
Such an approach we will apply to  many-valued predicate logic
(without quantifiers), with the set of k-ary predicate letters in
$P$, and to its particular
propositional case (where all predicate letters have 0-arity).\\
  This paper follows the following plan:\\ After a brief introduction to the truth-invariance  inference semantics, multi-modal \emph{predicate} logics (without quantifiers),
  and  a short introduction to binary sequents and bivaluations, in
  Section 2 we present the reduction of finite many-valued
  propositional (and predicate) logic language $\L$ into the 2-valued multi-modal
  algebraic logic language $\L_M$. Then we show the main properties for this positive logic (with standard
  2-valued conjunction and disjunction and a  modal operator $[x]$ for each  truth
  value $x$
  in a finite set $X$). We  present  a normal-forms reduction as well, where each obtained formula has the modal operators applied only to
  propositional letters in $\L$ (atoms in predicate logic with
  Herbrand base $H$). In Section 3 we develop the
  binary  sequent system $\G$, by extending the classic 2-valued
  distributive lattice logic (DLL) with the set of sequent axioms for
  each logic connective of a many-valued logic language $\L$. We show that
  this proof-theoretic sequent logic is sound and complete w.r.t. the
  model-theoretic semantics, based on  many-valued valuations:
  each \emph{deduced} sequent from $\G$ and a given set of sequent
  assumptions $\Gamma$ is also a \emph{valid} sequent (satisfied for every
  many-valued valuation), and viceversa.\\
  Finally in Section 4 we develop an autoreferential Kripke-style
  semantics,  based on Lindenbaum algebra of a many-valued
  logic language $\L$, and we define the  Kripke frame for it with the set of possible worlds equal to the set of
   truth values $X$.
  Than we show that this semantics is correct (sound and complete)
  for the multi-modal logic language $\L_M$, that is, we demonstrate that for each many-valued
  algebraic  model we obtain a correspondent Kripke model, so that a formula
  true in $\L_M$ is true  in this Kripke model as well, and
  viceversa.\\
  In what follows we  denote by  $B^A$ the set of all functions from
$A$ to $B$,  by $A^n$ a n-fold cartesian product $A \times ...\times
A$ for $n \geq 1$, and by $\P(A)$ the set of all subsets of $A$.
\subsection{Truth-invariance model-theoretic entailment} \label{subsec:truth-invariance}
In the standard 2-valued model-theoretic semantics we say that a
valuation $v:\L \rightarrow \textbf{2}$ is a \emph{model} of a
sentence $\psi \in \L$  iff $v(\psi) = 1$ (here $\L$ denotes the set
of all ground formulae of a given logic language). Consequently, a
formula $\phi$ is deduced from the set of formulae $\Gamma \subseteq
\L$, denoted by $\Gamma \models_1 \phi$, iff $~\forall v \in
Mod_{\Gamma}.(v(\phi) = 1)$, where $Mod_{\Gamma}$ is the set of all
models of the formulae in $\Gamma$ (here we use the index $1 \in
\textbf{2}$ in the consequence relation $\models_1$ to indicate the
deduction of the \emph{true}
formulae).\\
The set $\Gamma$ can be formally constructed by a subset of formulae
$\Gamma_1$ that we want to be (always) true, and by a subset of
formulae $\Gamma_0$ that we want to be (always) \emph{false}, so
that, $\Gamma = \Gamma_1 \bigcup \{\neg \phi~ |
\phi \in \Gamma_0 ~\}$ where $\neg$ is the 2-valued negation operator ($\neg 1 = 0, \neg 0 = 1$).\\
What is not often underlined is that this standard 2-valued
model-theoretic semantics \emph{implicitly} defines the set of
\emph{false} sentences deduced from $\Gamma$ as well, denoted here
explicitly by the new derived symbol $\models_0$ for the deduction
of false sentences (with the index $0 \in \textbf{2}$), by:
$~~~\Gamma \models_0 \phi~$ iff $~\Gamma \models_1 \neg\phi$, that
is,
iff $~\forall v \in Mod_{\Gamma}.(v(\phi) = 0)$.\\
Consequently, the classic 2-valued truth-invariance semantics of
logic entailment can be
paraphrased by the following generalized entailment, denoted by $\Gamma \models \phi$:\\
(CL)$~~~~$ "a formula $\phi$ is a logic consequence of the set
$\Gamma$"$~$ iff $~(\exists x \in
\textbf{2})(\forall v \in Mod_{\Gamma}).(v(\phi) = x)$.\\
 Thus,   classic 2-valued entailment deduces both true and false sentences if they have the
 same (i.e., \emph{invariant}) truth-value in all models of $\Gamma$.
 The consequence relation  $\models_1$ defines Tarskian closure
operator $C:\P(\L) \rightarrow \P(\L)$, such that $C(\Gamma) =
\{\phi~|~\phi \in \L$ and $ \Gamma \models_1 \phi\}$. In the
2-valued logics we do not need to use the consequence relation
$\models_0$ because the set of false sentences deduced from $\Gamma$
is equal to the set $\{\neg \phi~|~\phi \in \L$ and $ \Gamma
\models_1 \phi\} = \{\neg \phi~|~\phi \in C(\Gamma)\}$. This
particular property explains why in the classic 2-valued logic it is
enough to consider only the consequence relation for deduction of
true sentences, or alternatively the Tarskian closure operator $C$.
\\
But in the case of many-valued logics it is not generally the case,
and we need the consequence relations for the derivation of
sentences that are not true as well.  Consequently, this classic
2-valued model-theoretic truth-invariance semantics of logic
entailment we will
 directly extend to many-valued logics by: \\
 (MV) $~~~~\Gamma \models \phi~$ iff $~(\exists x \in
X)(\forall v \in Mod_{\Gamma}).(v(\phi) = x)$,\\
 where $Mod_{\Gamma}$ is specified by prefixing a particular truth-value $y
 \in X$ to each formula $\psi \in \Gamma$, as we explained previously in
 the case of the 2-valued logic with two subsets $\Gamma_1$ and
 $\Gamma_0$ for prefixed true and false sentences. \\
  The matrix-based inference is different and specified by:\\
(MX) $~~~~\Gamma \models \phi~$ iff $~(\forall v \in Mod_{\Gamma,D}).(v(\phi) \in D)$,\\
with the set of models $Mod_{\Gamma,D}$ of $\Gamma$ defined as the
set of
valuations $v$ such that $(\forall \psi \in \Gamma).(v(\psi) \in \D)$.\\
It is easy to verify that both semantics, the truth-invariance (MV)
and the matrix-based (MX), in the case of classic 2-valued logics,
where $D
= \{1\}$, coincide.\\
This new truth-invariance semantics of logic entailment for
many-valued logics has been presented first time in \cite{MajkC04}
and successively used for a new representation theorem for
many-valued modal logics  \cite{Majk06th}.

\subsection{Introduction to multi-modal predicate logic}
\label{subsection1}
%

More exhaustive and formal introduction to modal logics and their
Kripke models can be easily found in a literature, for example in
\cite{BBWo06}. Here will be given only a short informal version,
in order to make more clear definitions used in next paragraphs.\\
A predicate multi-modal logic, for a language with a set of
predicate symbols $r \in P$ with arity $ar(r)\geq 0$ and a set of
functional symbols $f \in F$ with arity $ar(f) \geq 0$, is a
standard predicate logic extended by a \emph{finite} number of
universal modal operators $\Box_i, i \geq 1$. In this case we do not
require that these universal modal operators are normal  (that is,
monotonic and multiplicative) modal operators as in the standard
setting for modal logics, but we require that they have the same
standard Kripke semantics. In the standard Kripke semantics each
modal operator $\Box_i$ is defined by an accessibility binary
relation $\R_i \subseteq \W \times \W$ for  a given set of possible worlds $\W$. \\
We define the set of terms of this predicate modal logic  as
follows:
 all variables $x \in Var$, and constants $d \in S$ are terms; if $f \in F$ is a functional symbol of arity $k = ar(f)$ and $t_1,..,
t_k$ are terms, then $~f(t_1,..,t_k)$ is a term. We denote by $\T_0$ the set of all ground (without variables) terms. \\
An atomic formula (atom) for a predicate symbol $r \in P$ with arity
$k = ar(r)$ is an expression $r(t_1,...,t_k)$, where $t_i, i
=1,...,k$ are terms. Herbrand base $H$ is a set of all ground atoms
(atoms without variables). More complex formulae, for a predicate
multi-modal logic, are obtained as a free algebra obtained from set
of all atoms and usual set of classic 2-valued binary logic
connectives in $ \{\wedge, \vee, \Rightarrow\}$ for conjunction,
disjunction and implication respectively (negation of a formula
$\phi$, denoted by $\neg \phi$ is expressed by $\phi \Rightarrow 0$,
where $0$ is used for an inconsistent formula (has constantly value
$0$ for every valuation), and  a number of unary universal modal
operators $\Box_i$. We define $\N = \{1,2,..., n\}$ where $n$ is the
maximal arity of symbols in the finite set $P \bigcup F$.
\begin{definition} \label{def:KripSem}
 We denote by $\M = (\W, \{$$
{\R}_i~ | ~1 \leq i \leq k \}, S, V)$ a multi-modal  Kripke model
with finite $k \geq 1$ modal operators with
 a set of possible worlds $\W$, the accessibility relations ${\R}_i \subseteq \W  \times \W$,
 non empty set of individuals $S$, and    a function $~~V:\W\times (P \bigcup F) \rightarrow {\bigcup}_{n \in \N}
(\textbf{2}\bigcup S)^{S^n}$, such that for any world $w \in \W$,\\
1. For any functional letter $f \in F$, $~V(w,f):S^{ar(f)}
\rightarrow S~$ is a
function (interpretation of $f$ in $w$).\\
2. For any predicate letter $r \in P$, the function
$~V(w,r):S^{ar(r)} \rightarrow \textbf{2}~$ defines the extension of
$r$ in a world $w$, $~~\|r\| = \{ \textbf{d}  = <d_1,...,d_k> \in
S^k~|~ k = ar(r), V(w,r)(\textbf{d}) = 1 \}$.
 \end{definition}
  We denote by   $~~{\M} \models_{w,g}~\varphi~$ the fact that a formula $~\varphi$ is satisfied in a world $w \in \W$ for
 a given assignment $g:Var \rightarrow S$. For example, a given atom $r(x_1,...,x_k)$ is satisfied in $w$ by assignment $g$,
 i.e., $~{\M} \models_{w,g}~r(x_1,...,x_k),~$ iff $~V(w,r)(g(x_1),...,g(x_k)) = 1$.\\
 The Kripke semantics is extended to all formuale as follows:
 \\$~~{\M} \models_{w,g}~ \varphi \wedge \phi~~~$ iff $~~~{\M} \models_{w,g}~ \varphi~$ and $~{\M} \models_{w,g}~
 \phi~$,
 \\$~~{\M} \models_{w,g}~ \varphi \vee \phi~~~$ iff $~~~{\M} \models_{w,g}~ \varphi~$ or $~{\M} \models_{w,g}~
 \phi~$,
 \\$~~{\M} \models_{w,g}~ \varphi \Rightarrow \phi~~~$ iff $~~~{\M} \models_{w,g}~ \varphi~$ implies $~{\M} \models_{w,g}~
 \phi~$,
\\ $~~{\M} \models_{w,g}~\Box_i \varphi~~~$ iff $~~~\forall w'((w,w')
\in {\R}_i $ implies ${\M} \models_{w',g}~ \varphi~)~$.\\ The
existential
modal operator $\diamondsuit_i $ can be defined as a derived operator by taking $\neg \Box_i \neg$.\\
A formula $\varphi$ is said to be \emph{true in a model} ${\M}$ if
 for each assignment function $g$
and possible world $w$, ${\M} \models_{w,g}~\varphi$. A formula is
said to be \emph{valid} if it
is true in each model.\\
We denote by $\|\phi/g\| = \{w~|~\M \} \models_{w,g}~\phi\}$ the set
of all worlds where the ground formula $\phi/g$ (obtained from
$\phi$ and an assignment $g$) is satisfied.\\
Remark: in this paper we will use the notation $[x]$ (for any truth
value $x \in X$) for universal modal operators, instead of standard
notation $\Box_i$.
\subsection{Introduction to binary sequents and bivaluations}
Sequent calculus has been developed by Gentzen \cite{Gent32},
inspired on ideas of Paul Hertz \cite{Hert29}. Given a propositional
logic language $\L_A$ (a set of logic formulae) a sequent
 is a consequence pair of formulae, $ s = (\phi ; \psi) \in \L_A \times \L_A$, denoted also by $\phi \vdash
 \psi$. \\A Gentzen system, denoted by a pair $~\G = \langle \mathbb{L}, \Vdash
 \rangle$ where $\Vdash$ is a consequence relation on a set of sequents in $\mathbb{L} \subseteq \L_A \times
 \L_A$, is said  \emph{normal} if it satisfies the following
 conditions: for any sequent $s = \phi \vdash
 \psi \in \mathbb{L}$ and a set of
 sequents $\Gamma = \{s_i = \phi_i \vdash \psi_i \in \mathbb{L}~|~i \in I\}$,\\
 1. (reflexivity) if $s \in \Gamma$ then $\Gamma \Vdash s$\\
 2. (transitivity) if $\Gamma \Vdash s$ and for every $s' \in
 \Gamma$, $\Theta \Vdash s'$, then $\Theta \Vdash s$\\
 3. (finiteness) if $\Gamma \Vdash s$ then there is finite $\Theta
 \subseteq\Gamma$ such that $\Theta \Vdash s$.\\
 4. for any homomorphism $\sigma$ from $\mathbb{L}$ into itself
 (i.e., a substitution), if $\Gamma \Vdash s$ then $\sigma[\Gamma] \Vdash
 \sigma(s)$, i.e., $\{\sigma(\phi_i) \vdash \sigma(\psi_i)~|~i \in I\} \Vdash (\sigma(\phi) \vdash \sigma(\psi))$.\\
 Notice that from (1) and (2) we obtain this monotonic property:\\
 5.  if  $\Gamma \Vdash s$ and $\Gamma
 \subseteq\Theta$, then $\Theta \Vdash s$.\\
 We denote by $C:\P(\mathbb{L})\rightarrow \P(\mathbb{L})$ the Tarskian
 closure  operator such that $C(\Gamma) =_{def} \{ s \in \mathbb{L}~|~\Gamma \Vdash
 s\}$, with the properties: $\Gamma \subseteq C(\Gamma)$ (from
 reflexivity (1)); it is monotonic, i.e., $\Gamma \subseteq \Gamma_1$ implies
 $C(\Gamma) \subseteq C(\Gamma_1)$ (from  (5)), and an
 involution  $C(C(\Gamma)) = C(\Gamma)$ as well. Thus, we obtain that\\
 6. $\Gamma \Vdash s ~$ iff $~s \in C(\Gamma)$.\\
 Any sequent theory $\Gamma \subseteq \mathbb{L}$ is said  a \emph{closed}
 theory iff $\Gamma = C(\Gamma)$. This closure property corresponds to the
 fact that $\Gamma \Vdash s$ iff $s \in \Gamma$.\\
 Each sequent theory $\Gamma$ can be considered as a bivaluation (a characteristic function) $\beta:\mathbb{L} \rightarrow
 \textbf{2}$ such that for any sequent $s \in \mathbb{L}$, $~\beta(s)
 = 1~$ iff $~s \in \Gamma$.
%
%
\section{Reduction of finite many-valued into 2-valued multi-modal
logic} \label{section2}
In what follows let $\L_P$ be a predicate  logic language obtained
as a free algebra, from connectives  in $\Sigma$ of an algebra with
a set $X$ of  truth values (for example the many-valued conjunction,
disjunction and implication $\{\wedge_m, \vee_m, \Rightarrow_m \}
\subseteq \Sigma$ are binary operators, negation $\neg_m \in \Sigma$
and other modal operators are unary operators, while each $x \in X
\subseteq \Sigma$ is a constant (nullary operator)), a set $P$ of
predicate symbols   denoted by $p,r,q,..$ with a given arity (in the
case when the arity of all symbols in $P$ is a zero we obtain that
$P$ is the set of propositional variables (letters), so that $\L_P$
is a propositional logic), and a set $F$ of functional symbols (with
a given arity) denoted by $f,g,h.$.\\ We define the set of terms of
this logic as follows:
 all variables $\nu_i \in Var$, $i = 1,2,...$, and constants $d \in S$ are terms; if $f \in F$ is a functional symbol of arity $n$ and $t_1,..,
t_n$ are terms, then $~f(t_1,..,t_n)$ is a term. The ground term is
a term without variables. We denote by $\T_S$ a set of all terms.\\
 The set of atoms is
defined as $~~\A_S = \{p(c_1,..,c_n)~|~ p \in P,~n = arity(p)~$ and
$~ c_i \in \T_S\}$.  The set of all ground atoms (without
variables), $~~H = \{p(c_1/g,..,c_n/g)~|~ p \in P,~n = arity(p)~$,
$~ c_i
\in \T_S$ and $g:Var \rightarrow S\}$, is  a Herbrand base (here $c_i/g$ denotes a ground atom obtained from a term $c_i$ by an assignment $g$).\\
 Any atom is a logic formulae. The combination of logic formulae by
logical connectives in $\Sigma$ is another logic formula.\\
We will denote by $\L$ the subset of the logic language $\L_P$
composed of only \emph{ground} formulae. In this case we can
nominate each ground atom by a particular propositional letter, so
that this logic language $\L$ is equivalent to the propositional
logic language where the ground atoms in $H \subseteq \L$ are
replaced by propositional letters $A,B,..$.\\
 We will use the letters $\phi, \psi$
for formulae of $\L$.\\
We define a (many-valued) \emph{valuation} $v$ as a mapping $v: H
\rightarrow X$,
which is uniquely extended in standard way to the  homomorphism $v:
\L \rightarrow X$ (for example,
 for any $A,B \in H$, $v(A \odot B) = v(A) \odot v(B), \odot \in \{\wedge_m, \vee_m, \Rightarrow_m \}$ and $v(\neg_m A) = \neg_m v(A)$, where
 $\wedge_m, \vee_m, \Rightarrow_m, \neg_m$ are the many-valued conjunction, disjunction, implication and negation
 respectively). The set of all \emph{many-valued valuations} is a strict subset $\mathbb{V}_m$ of the functional space $X^{\L}$
 that satisfies the homomorphic conditions above.\\
Based on this propositional many-valued logic language $\L$ we are
able to define the following multi-modal 2-valued algebraic logic
language $\L^*_{M}$, by introduction \cite{Majk06MV} of modal
non-standard (non monotonic) algebraic truth-functional operators
$[x]: Y \rightarrow \textbf{2}$, where $Y = X \bigcup \textbf{2}$,
such that for any $x, y  \in Y$, $~~[x](y) = 1~~$ iff $~~x = y$.
The intersection of $X$ and $\textbf{2}$ can be non empty as well.\\
The set of  truth values $X$ is a finite set, so that the number of
these algebraic modal operators $n = |Y|\geq 2 $ is finite as well
($|Y|$ is the cardinality of the set $Y$).
\begin{definition}
 \label{def:multimodal} \textsc{Syntax:} Let $\L_P$ be a predicate many-valued
 logic language with a set of  truth values $X$ and $(\textbf{2}, \wedge,
 \vee)$ the complete distributive two-valued lattice.
 The  multi-modal 2-valued logic language $\L_{M}^*$, is
 the set of all modal formulae (we will use
 letters $\Phi,\Psi..$ for the formulae of $\L_{M}^*$) defined as follows:\\
1. $~~\textbf{2} \subseteq \L_{M}^*$.\\
2. $~~[x] \phi \in \L_{M}^*$, for any $x \in X, \phi \in \L_P$.\\
3. $~~[x] \Phi \in \L_{M}^*$, for any $x \in  \textbf{2}, \Phi \in \L_M^*$.\\
4. $~~\Phi, \Psi \in \L_{M}^*$ implies $\Phi \wedge \Psi, \Phi \vee
\Psi \in \L_{M}^*$.
\\ We denote by $\L_M$ the  sublanguage
of $\L_M^*$ without variables (ground atoms in $\L_M$ are considered
as propositional letters).
\end{definition}
 The constants $0,1$ correspond to the tautology and contradiction
 proposition respectively, and can be considered as nullary operators in
 $\L_{M}$.
\\ We can use this 2-valued multi-modal logic language $\L_{M}$ in
 order to define the sequents as elements of the cartesian product
 $\L_{M}\times \L_{M}$, i.e., each sequent $s$, denoted by $\Phi
 \vdash \Psi$, where $\Phi, \Psi \in \L_{M}$.
\begin{definition} \label{def:mvaluation} \textsc{Semantics:}
For any many-valued valuation  $v \in \mathbb{V}_m$,  $v:\L
\rightarrow
 X$, we define the  'modal valuation' $~\alpha :\L_M \rightarrow \textbf{2}$ as follows: \\
1. $~~\alpha(0) = 0$, $~~\alpha(1) = 1$.\\
2. $~~\alpha([x]\phi) = 1~~$ iff $~~x = v(\phi)$, for any $x \in X$, $\phi \in \L$.\\
 3. $~~\alpha([x]\Phi) = 1~~$ iff $~~x = \alpha(\Phi)$, for any $x \in \textbf{2}$, $\Phi \in \L_{M}$. \\
4. $~~\alpha(\Phi \wedge \Psi) = \alpha(\Phi) \wedge \alpha(\Psi)$,
$~~\alpha(\Phi \vee \Psi) = \alpha(\Phi) \vee \alpha(\Psi)$, for any
$~\Phi, \Psi \in \L_{M}$.
\\ This transformation from many-valued valuations into modal valuatins
can be expressed by the mapping $~\mathfrak{F}:\mathbb{V}_m
\rightarrow  \mathcal{V}$, where $\V \subset \textbf{2}^{\L_{M}}$
denotes the set of all modal valuations.
\end{definition}
It is easy to verify that the mapping $~\mathfrak{F}$ is a
 bijection, with its inverse $~\mathfrak{F}^{-1}$ defined as
 follows: for any modal valuation $\alpha \in \V$,  given in Definition
 \ref{def:mvaluation}, we have that the many-valued valuation $v
 = \mathfrak{F}^{-1}(\alpha):\L \rightarrow X$ is defined  for any $\phi \in \L$, by
 $~v(\phi) = x \in X~$ iff $~\alpha([x]\phi) = 1$.\\
 A many-valued valuation $v:\L \rightarrow X$, $v \in \mathbb{V}_m$,  satisfies a 2-valued multi-modal formula $\Phi \in
 \L_{M}~~$ iff $~~\mathfrak{F}(v)(\Phi) =1$.\\
 Given two formulae $\Phi, \Psi \in \L_M$, the sequent $\Phi \vdash
\Psi$ is satisfied by $~v~$ if $~~\mathfrak{F}(v)(\Phi) \leq
 ~\mathfrak{F}(v)(\Psi)$. A sequent $\Phi \vdash \Psi$ is an axiom
 if it is satisfied by every valuation $v \in \mathbb{V}_m\subset X^{\L}$.\\
From this definition of satisfaction for sequents we obtain the
 reflexivity (axiom) $\Phi \vdash \Phi$ and transitivity (cut)
 inference rule, i.e., from $\Phi \vdash \Psi$ and $\Psi \vdash
 \Upsilon$ we deduce $\Phi \vdash \Upsilon$.
\\ Let us define the  set of  2-valued multi-modal literals (or modal atoms),\\ $P_{mm} = \{[x_1]...[x_k]A \in \L_M~|~k \geq 1$ and   $A \in H \}$.\\
 For example, if $v(\phi) = x$ then $~\mathfrak{F}(v)([1][x]\phi) = 1$, while if
$v(\phi) \neq x$ then $~\mathfrak{F}(v)([0][x]\phi)\\ = 1$. Notice
that the number of nested modal operators can be reduced from the
fact that $[0][0]$ and $[1][1]$ are identities for the formulae in
$\L_M$. For example, for $~[x]\phi \in \L_M$, we have
$~[0][1][1][0][x]\phi
\equiv [0][0][x]\phi \equiv [x]\phi$, where $~\equiv~$ is a standard logic equivalence. \\
 Then, given a formula $\phi \in \L$,  we have that
the modal formula $[x]\phi \in \L_{M}$ can be naturally reduced into
an equivalent formula, denoted by $\widehat{[x]\phi}$, where the
modal operators $[x]$ are applied only to ground atoms (considered
as propositional letters) in $H$. Moreover, for any formula $\Phi
\in \L_M$ there is an equivalent formula $\widehat{\Phi}$ composed
by logical connectives $\wedge$, $\vee$, and by multi-modal literals
in $P_{mm}$.  A canonical formula can be obtained by the following
reduction:
\begin{definition} \textsc{Canonical reduction:} \label{def:reduction} Let us define the following reduction rules:
1. For any unary operator $~\sim \in \Sigma$, $~\phi \in \L$, and a value $x \in X$,\\
$[x](\sim \phi)~~~~ \mapsto ~~~~ \bigvee_{y\in X. ~x = \sim y}
[y]\phi$,\\
2. For any binary operator $~\odot \in \Sigma$,  $~\phi, \psi \in \L$, and a value $x \in X$,\\
$[x]( \phi \odot \psi )~~~~ \mapsto ~~~~ \bigvee_{y,z\in X. ~x =
y\odot z} ([y]\phi \wedge [z]\psi)$.\\
3. For any binary operator $~\odot \in \{\wedge, \vee \}$,  $~\Phi, \Psi \in \L_M$, and a value $x \in  \textbf{2}$,\\
$[x]( \Phi \odot \Psi )~~~~ \mapsto ~~~~ \bigvee_{y,z\in
~\textbf{2}.~ x = y\odot z} ([y]\Phi \wedge [z]\Psi)$.
\\We denote by $~\widehat{[x]\Phi}~$ the  canonic formula obtained  by applying recursively these
reduction rules to the formula $~[x]\Phi$.
\end{definition}
\begin{propo} \label{prop:reduction}
The normal reductions in Definition \ref{def:reduction} are
truth-preserving, that is, for any $~x \in X$ and $~\phi \in \L$ we
have that  $~~[x]\phi~$ is logically equivalent to $~\widehat{[x]\phi}$.\\
Analogously, for any $~x \in \textbf{2}$ and $~\Phi \in \L_M$ we
have that $~~[x]\Phi~$ is logically equivalent to
$~\widehat{[x]\Phi}$.
 \end{propo}
\textbf{Proof:} Let us show that the steps of canonical reduction in Definition \ref{def:reduction} are truth-preserving:\\
1. The first case of reduction: let us suppose that $[x](\sim \phi)$
is true but $\bigvee_{y\in X. x = \sim y} [y]\phi$ is false. From
the truth of $[x](\sim \phi)$ we obtain that $x = v(\sim \phi)$: let
$z = v(\phi)$ and, consequently, $[z]\phi$ is true, then, from the
truth-functional connective $\sim$ we conclude that $x = \sim z$ and
from the fact that $[z]\phi$ is true, we conclude that
$\bigvee_{y\in X. x = \sim y} [y]\phi$ must be true, which is a
contradiction. Consequently $[x](\sim \phi)$ implies $\bigvee_{y\in
X. x = \sim y} [y]\phi$.\\
Viceversa, if $\bigvee_{y\in X. x = \sim y} [y]\phi$ is true then
there exists $z$ such that $[z]\phi$ is true, with $ x =\sim z$,
i.e., $x
= v(\sim \phi)$ so that $[x](\sim \phi)$ is true.\\
Consequently,  $[x](\sim \phi)~~$ iff $~~\bigvee_{y\in X. x = \sim
y} [y]\phi$. \\
2. Analogously, for the second case (and 3rd as well) we obtain that
\\$[x]( \phi \odot \psi )~~$ iff $~~ \bigvee_{y\in X. x = y\odot z}
([y]\phi \wedge [z]\psi)$.\\
From the fact that both steps are also truth-preserving, we deduce
that any consecutive execution of them is truth-preserving, and,
consequently, $~[x]\phi~~$ iff $~~\widehat{[x]\phi}$.
\\$\square$\\
The following proposition shows that the result of the canonical
reduction of a formula $~\Phi \in \L_M$ is a  disjunction of modal
conjunctions, which in the case of the formulae without nested modal
operators is a simple disjunctive modal formula.
\begin{propo} \label{prop:norform} Any 2-valued logic formulae
 $~\Phi \in \L_M$ is logically equivalent to  disjunctive modal
 formula $~ \bigvee_{1 \leq i \leq
 m}(\bigwedge_{1 \leq j \leq m_i}
 ([y_{ij1}]...[y_{ijk_{ij}}])A_{ij})$, where for all $1 \leq i \leq
 m$,  and $1 \leq j \leq m_i$, we have that $1\leq k_{ij}$, $~[y_{ijk_{ij}}] \in X$,  and  $~A_{ij} \in H$.\\ In the case when we have no
 nested modal operators then $k_{ij} = 1$ for all $i,j$. Consequently,  $~\Phi \in \L_M$ is logically equivalent to a disjunctive modal
 formula
 $~\bigvee_{1 \leq i \leq m} [x_i] \phi_i$, where for all $1 \leq i \leq
 m$, $~x_i \in X$, $~\phi_i \in \L$.
 \end{propo}
 \textbf{Proof:} From  Proposition \ref{prop:reduction} and from the canonical reduction in Definition \ref{def:reduction}, it is easy to
 see that each logically equivalent reduction moves a modal operator toward propositional letters in $H$, by introduction of
  conjunction and disjunction logic operators only. Thus, when this
  reduction is completely realized  we obtain a positive propositional logic
  formula with modal propositions in $P_{mm}$ and logic operators
  $\wedge$ and $\vee$. It is well known that such a positive propositional
  formula can be equivalently (but not uniquely) represented as a disjunction of
  conjunctions of modal propositions
  $([y_{ij1}]...[y_{ijk_{ij}}])A_{ij}  \in P_{mm}$.\\
 Let us consider now a formula $~\Phi \in \L_M$ without nested modal operators. Then  $k_{ij} = 1$ for all $i,j$
 and,  from the result above, a formula $\Phi \in \L_M$
 can be equivalently (but not uniquely) represented as a disjunction of conjunctive forms  $~ \bigvee_{1 \leq i \leq
 m}(\bigwedge_{1 \leq j \leq m_i} [y_{ij1}]A_{ij})$.\\
 But, we have that $~\bigwedge_{1 \leq j \leq m_i} [y_{ij1}]A_{ij} =
 [x_i]\phi_i$, where for any binary operator $\odot \in \Sigma$,
 $~x_i = y_{i11}\odot ...\odot y_{im_i1}~ \in X$, and $~\phi_i = A_{i1}\odot ...\odot A_{im_i}~ \in
 \L$.
 It holds because $\alpha([x_i]\phi_i) = 1~$ iff $~x_i = v(\phi_i) = v(A_{i1}\odot ...\odot
 A_{im_i}) = v(A_{i1})\odot ...\odot v(A_{im_i}) = y_{i11}\odot ...\odot
 y_{im_i1}$.\\
 That is, we obtain that holds:  $~\Phi ~$ iff $~ \bigvee_{1 \leq i \leq m}[x_i] \phi_i$.
 \\$\square$\\
 With this normal reduction, by using truth-value tables of
many-valued logical connectives, we  introduced  the structural
compositionality and  truth preserving for the 2-valued modal
encapsulation of a many-valued logic $\L$ as well. In fact, the
following property is valid:
\begin{propo}  \label{prop:preserving}
 Given a many-valued valuation $v:\L \rightarrow X$ and a formula
 $\phi \in \L$, the normal reduct formula $~\widehat{[x]\phi} \in
 \L_{M}$ is satisfied by $~v~~$ iff $~~x = v(\phi)$.
 \end{propo}
\textbf{ Proof:} By structural induction on $\phi$: let $\alpha = \mathfrak{F}(v)$, then\\
1. Case when $\phi = A \in H$:\\ If $~\widehat{[x]A}$ is satisfied
than $\alpha(\widehat{[x]\phi}) = \alpha([x]A) = 1$ (from
Definition \ref{def:mvaluation} and Proposition \ref{prop:reduction}), thus $x = v(A) = v(\phi)$.\\
2. Case when $\phi = \sim \psi$, where $\sim$ is an unary operator
in $\Sigma$, with $y = v(\psi)$ and $x = \sim y$:\\ Suppose, by
structural induction, that $\widehat{[y]\psi}$ is satisfied, i.e.,
$y = v(\psi)$, so that $[y]\psi$ is true (from Definition
\ref{def:mvaluation}). Thus, $\bigvee_{y\in X. x = \sim y} [y]\psi$
is true  and, by the truth-preservation (from the reduction 1 in
Definition \ref{def:reduction}) we obtain that $[x](\sim \psi)$ is
true, so that, $ [x] \phi$ is true, i.e., (by Definition
\ref{def:mvaluation}) $x = v(\phi)$, and (from Proposition
\ref{prop:reduction}) $ \widehat{[x] \phi}$ is true.
 With $x = \sim y = \sim
(v(\psi)) =$ (from the homomorphism of $v$) $
= v (\sim \psi) = v(\phi)$.\\
3.  Case when $\phi =  \psi\odot \varphi$, where $\odot$ is a binary
operator in $\Sigma$, with $y = v(\psi)$, $z = v(\varphi)$ and $x =
 y \odot z$:\\
  Then, by structural induction hypothesis, $\widehat{[y]\psi}$ and  $\widehat{[z]\varphi}$
 are
satisfied, so that from Proposition \ref{prop:reduction} and
Definition \ref{def:mvaluation} we have that $~1 =
\alpha(\widehat{[y]\psi}) = \alpha([y]\psi)$ and $1 =
\alpha(\widehat{[z]\varphi}) = \alpha([z]\varphi)$. Consequently, $1
= \alpha([y]\psi) \wedge \alpha([z]\varphi) = $ (from the
homomorphism 2 in Definition \ref{def:mvaluation}) $ = \alpha
([y]\psi \wedge [z]\varphi) = \alpha(\bigvee_{y\in X. x = y\odot z}
([y]\psi \wedge [z]\varphi)) =$ (from the reduction 2 in Definition
\ref{def:reduction}) $ = \alpha([x](\psi \odot \varphi)) =
\alpha([x]\phi) =$ (by Proposition \ref{prop:reduction}) $ =
\alpha(\widehat{[x]\phi})$. Thus, $~\widehat{[x]\phi}$ is satisfied
by $v$ and $x =  y \odot z = v(\psi) \odot v(\varphi) =$ (from the
homomorphism of $v$) $ = v ( \psi \odot \varphi) = v(\phi)$.
\\$\square$\\
Thus, as a consequence, for any $\phi \in \L$ and a many-valued
valuation $v \in \mathbb{V}_m$,  we have that
$~~\mathfrak{F}(v)(\widehat{[x]\phi}) = 1~~$ iff $~~x = v(\phi)$.

\section{Many-valued truth and model theoretic  semantics}
 The Gentzen-like system $\G $ of the 2-valued propositional logic  $\L_{M}$ (where the set of propositional letters corresponds to the
 set $P_{mm} = \{[x_1]...[x_k]A \in \L_M~|~k \geq 1$ and $A \in H \}$,
 is a 2-valued distributive logic  (DLL in \cite{Dunn95}),
  that is, $\textbf{2} \subseteq \L_{M}$, extended by
 the set of sequent axioms, defined for each many-valued logic connective in $\Sigma$ of the original many-valued logic $\L$. It contains the
 following axioms (sequents) and rules: \\\\
 (AXIOMS) The Gentzen-like system $\G = \langle \mathbb{L}, \Vdash
 \rangle$ contains  the following sequents in $\mathbb{L}$, for any $\Phi, \Psi, \Upsilon \in
 \L_{M}$:\\
 1. $~\Phi \vdash \Phi~~$ (reflexive)\\
 2. $~\Phi \vdash 1$, $~~0 \vdash \Phi~~$ (top/bottom axioms)\\
  3. $~\Phi \wedge \Psi \vdash \Phi$, $~~\Phi \wedge \Psi \vdash \Psi~~$ (product projections: axioms for
 meet)\\
 4. $~\Phi  \vdash \Phi \vee \Psi$, $~~\Phi \vdash \Psi \vee \Phi~~$ (coproduct injections: axioms
 for join)\\
 5. $~\Phi \wedge (\Psi \vee \Upsilon) \vdash (\Phi \wedge \Psi) \vee (\Phi \wedge \Upsilon)~~$
 (distributivity axiom)\\
 6. The set of Introduction axioms for many-valued connectives:\\
 6.1 $~\bigvee_{y,z\in
~\textbf{2}.~ x_1 = y\odot z} ([y]\Phi \wedge [z]\Psi)~~ \vdash ~~ [x_1](\Phi \odot \Psi )$, for any binary operator $~\odot \in \{\wedge, \vee \}$.\\
 6.2 $~\bigvee_{y\in X. x = \sim y} [y]\phi~~ \vdash ~~ [x](\sim \phi)$, for any unary operator $\sim \in \Sigma$ and $\phi \in \L$. \\
6.3 $~\bigvee_{y,z\in X. x = y\odot z} ([y]\phi \wedge [z]\psi)~~
\vdash ~~ [x]( \phi \odot \psi )$, for any binary operator $\odot \in \Sigma$ and $\phi, \psi \in \L$.\\
 7. The set of Elimintion axioms for many-valued connectives:\\
7.1 $~[x_1]( \Phi \odot \Psi )~~ \vdash ~~ \bigvee_{y,z\in
~\textbf{2}.~ x_1 = y\odot z} ([y]\Phi \wedge [z]\Psi)$, for any binary operator $~\odot \in \{\wedge, \vee \}$.\\
 7.2 $~[x](\sim \phi)~~ \vdash ~~ \bigvee_{y\in X. x = \sim y}
[y]\phi$, for any unary operator $\sim \in \Sigma$ and $\phi \in \L$.\\
7.3 $~[x]( \phi \odot \psi )~~ \vdash ~~ \bigvee_{y,z\in X. x =
y\odot z}([y]\phi \wedge [z]\psi)$, for any binary operator $\odot \in \Sigma$ and $\phi, \psi \in \L$.\\\\
 (INFERENCE RULES) $\G$ is closed under the following inference
 rules:\\
 1. $~~\frac{\Phi ~\vdash \Psi,~~\Psi~ \vdash \Upsilon }{\Phi ~\vdash \Upsilon}~~$ (cut/
 transitivity rule)\\
 2. $~~\frac{\Phi ~\vdash \Psi,~~\Phi~ \vdash \Upsilon}{\Phi~ \vdash \Psi \wedge
 \Upsilon}~~$, $~~\frac{\Phi ~\vdash \Psi,~~\Upsilon~ \vdash \Psi}{\Phi \vee \Upsilon~ \vdash \Psi
 }~~$ (lower/upper lattice bound rules)\\
  3. $~~\frac{\Phi~\vdash \Psi}{\sigma(\Phi) ~ \vdash \sigma(\Psi)} ~~$ (substitution
 rule: $\sigma$ is substitution $(\gamma/p)$).
  \\$\square$ \\
 The axioms from 1 to 5 and the rules  1 and 2 are
 taken from \cite{Dunn95} for the $DLL$ and it was shown that this sequent based Genzen-like system is sound and
 complete.  The new axioms in points 6 and 7 corresponds to the canonical (equivalent) reductions in Definition \ref{def:reduction}. The set of sequents that define the poset of the classic 2-valued lattice of
 truth values $(\textbf{2}, \leq)$ is a consequence of the top/bottom axioms: for any two $x,y \in \textbf{2}$, if $x \leq y$
 then $x \vdash y \in \G$. \\
 Thus, for  many-valued logics we obtain a \emph{normal} modal Gentzen-like deductive system, where each
 sequent is a valid truth-preserving consequence-pair defined by the poset
 of the complete lattice $(\textbf{2}, \leq)$ of classic truth values (which are also
 the
 constants of this positive propositional language $\L_M$), so that each
 occurrence of the symbol $\vdash$ can be substituted by the partial
 order $\leq$ of this complete lattice $(\textbf{2}, \leq)$ .\\\\
 \textbf{Example 1:} Let us consider the Godel's 3-valued logic $X = \{0, \frac{1}{2}, 1 \}$, and
 its 3-valued implication logic connective $\Rightarrow$ given by
 the following truth-table:
 \begin{center}
\begin{tabular}{c|ccc}
  $~~\Rightarrow~~$ & $~~0~~$ & $~~\frac{1}{2}~~$ & $~~1~~$ \\
  \hline
  0 & 1 & 1 & 1 \\
    $\frac{1}{2}$ & 0 & 1 & 1 \\
  1 & 0 & $\frac{1}{2}$ & 1 \\
\end{tabular}
\end{center}
One of the possible m-sequents for introduction \emph{rules} for
this connective, taken from \cite{BFZa93}, (each rule corresponds to
the conjunction of disjunctive
forms, where each disjunctive form is one m-sequent in the premise), is\\
\begin{center}
$~~\frac{\langle \Gamma | \triangle, \phi |\Pi, \phi \rangle
~~\langle \Gamma_1, \psi| \triangle_1| \Pi_1\rangle}{\langle \Gamma,
\Gamma_1, \phi \Rightarrow \psi | \triangle, \triangle_1 | \Pi,
\Pi_1\rangle}~~\Rightarrow:0$, $~~\frac{\langle \Gamma | \triangle
|\Pi, \phi \rangle ~~\langle \Gamma_1| \triangle_1, \psi|
\Pi_1\rangle}{\langle \Gamma, \Gamma_1,
 | \triangle, \triangle_1, \phi \Rightarrow \psi | \Pi,
\Pi_1\rangle}~~\Rightarrow:\frac{1}{2}$\\
$~~\frac{\langle \Gamma, \phi | \triangle, \phi |\Pi, \psi \rangle
~~\langle \Gamma_1, \phi| \triangle_1, \psi| \Pi_1, \psi
\rangle}{\langle \Gamma, \Gamma_1,
 | \triangle, \triangle_1 | \Pi,
\Pi_1, \phi \Rightarrow \psi\rangle}~~\Rightarrow:1$
 \end{center}
 while in our approach we obtain the \emph{unique} set of binary sequent
 introduction \emph{axioms}:\\\\
 $~~([\frac{1}{2}]\phi \wedge [0]\psi) \vee ([1]\phi \wedge [0]\psi)
 ~~\vdash ~~[0](\phi \Rightarrow \psi)$\\
$~~[1]\phi \wedge [\frac{1}{2}]\psi ~~\vdash ~~[\frac{1}{2}](\phi \Rightarrow \psi)$\\
$~~\bigvee_{x,y \in X ~\&~ (x,y) \notin ~S}~([x]\phi \wedge [y]\psi) ~~\vdash ~~[1](\phi \Rightarrow \psi)$\\
where $~S =\{(\frac{1}{2},0), (1,0),(1,\frac{1}{2}) \}$, \\and
elimination axioms:\\
 $~~[0](\phi \Rightarrow \psi)
 ~~\vdash ~~([\frac{1}{2}]\phi \wedge [0]\psi) \vee ([1]\phi \wedge [0]\psi)$\\
$~~[\frac{1}{2}](\phi \Rightarrow \psi)~~\vdash ~~[1]\phi \wedge [\frac{1}{2}]\psi $\\
$~~[1](\phi \Rightarrow \psi) ~~\vdash ~~\bigvee_{x,y \in X ~\&~
(x,y) \notin ~S}~([x]\phi \wedge [y]\psi)$.
\\$\square$
 \begin{definition}
 \label{def:entailment} For any two formulae $\Phi, \Psi \in \L_M$
 when the
 sequent $\Phi \vdash \Psi$ is satisfied by a 2-valued modal valuation
  $\alpha:\L_M \rightarrow \textbf{2}$ from Definition \ref{def:mvaluation} (that is, when $~~\alpha(\Phi) \leq
 \alpha(\Psi)$ as in standard 2-valued logics), we say that it is satisfied by the many-valued
 valuation $v = \mathfrak{F}^{-1}(\alpha):\L\rightarrow X$.\\
 This sequent is a
 tautology if it is satisfied by all modal valuations $\alpha \in \V$, i.e., when
 $~\forall v \in \mathbb{V}_m .(\mathfrak{F}(v)(\Phi) \leq \mathfrak{F}(v)(\Psi))$.\\
 For a  normal Gentzen-like sequent system $\G = \langle \mathbb{L}, \Vdash
 \rangle$ of a many-valued logic language $\L$, with the set of sequents $\mathbb{L} \subseteq
 \L_M \times \L_M$, we tell that a many-valued valuation $v$ is its \verb"model"
  if it satisfies all sequents in $\G$. \\The set of all models of
  a given set of sequents (theory) $\Gamma$ is:\\
  $~~Mod_{\Gamma} = \{v\in \mathbb{V}_m~|~\forall(\Phi \vdash \Psi) \in \Gamma(\mathfrak{F}(v)(\Phi) \leq \mathfrak{F}(v)(\Psi))
  \} ~~\subseteq~  \mathbb{V}_m  \subset X^{\L}$.
 \end{definition}
 \begin{propo} \textsc{Sequent's bivaluations and soundness:} \label{prop:sound}
 Let us define the mapping $~\mathfrak{B}:\mathbb{V}_m\rightarrow \textbf{2}^{\L_M \times
\L_M}$ from valuations into sequent bivaluations, such that for any
valuation $~ v \in \mathbb{V}_m$ we obtain   the sequent bivaluation
$\beta = \mathfrak{B}(v) = eq \circ <\pi_1, \wedge
>\circ(\mathfrak{F}(v) \times \mathfrak{F}(v)):\L_M \times \L_M \rightarrow
\textbf{2}$,
 where $\pi_1$ is the first projection, $\circ$ is the
functional composition and $~eq:\textbf{2}\times
\textbf{2}\rightarrow \textbf{2}$ is the equality mapping such that
$~eq(x,y) = 1~$ iff $~x =y$. \\Than,  a sequent $~s =(\Phi
\vdash \Psi)$ is satisfied by $~v~$ iff $~\beta(s) = \mathfrak{B}(v)(s) = 1$.\\
 All  axioms of the Gentzen like sequent system $\G $, of a many-valued logic language $\L$ based on a set $X$ of  truth values,
  are the tautologies, and all its rules are sound for model satisfiability and preserve the  tautologies.
 \end{propo}
 \textbf{Proof:}
From the definition of a bivaluation $\beta$ we have that
\\$\beta(\Phi \vdash \Psi) = \beta (\Phi ; \Psi) = eq \circ <\pi_1,
\wedge
>\circ(\mathfrak{F}(v) \times \mathfrak{F}(v))(\Phi ; \Psi) \\= eq
\circ <\pi_1, \wedge >(\mathfrak{F}(v)(\Phi) \times
\mathfrak{F}(v)(\Psi)) \\~=~ eq <\pi_1(\mathfrak{F}(v)(\Phi) ,
\mathfrak{F}(v)(\Psi)), ~\wedge(\mathfrak{F}(v)(\Phi) ,
\mathfrak{F}(v)(\Psi))
> \\ =~ eq(\mathfrak{F}(v)(\Phi), ~\mathfrak{F}(v)(\Phi)
\wedge \mathfrak{F}(v)(\Psi))$.\\
 Thus $\beta(\Phi \vdash \Psi) = 1~~
$ iff $~~\mathfrak{F}(v)(\Phi) \leq \mathfrak{F}(v)(\Psi)$, i.e.,
 when this sequent is satisfied by $v$.\\
 It is straightforward to check  that all axioms in $\G$ are tautologies (all constant sequents
  specify the poset of the complete lattice $(\textbf{2}, \leq)$ of classic 2-valued logic, thus are tautologies).
  It is straightforward to check that all rules preserve the tautologies.
  Moreover, if all premises  of a given rule in $\G$ are satisfied by a given many-valued valuation $v:\L\rightarrow X$,
  then also the deduced sequent of this rule is satisfied by the same valuation, i.e., the rules are sound for the model satisfiability.
  \\$\square$ \\
  It is easy to verify, that this entailment is equal to the classic
 propositional entailment.\\
 \textbf{Remark:} It is easy to observe that each sequent is, from
 the logic point of view, a \\ \emph{2-valued object} so that all inference
 rules are embedded into the classic 2-valued framework, i.e., given a bivaluation $\beta = \mathfrak{B}(v):\L_M \times \L_M \rightarrow \textbf{2}$,
 we have that a sequent $s = \Phi \vdash \Psi$ is satisfied, $\beta(s) =
 1~~$ iff $~~\mathfrak{F}(v)(\Phi) \leq \mathfrak{F}(v)(\Psi)$,
  so that we have the direct relationship between sequent bivaluations $\beta$ and many-valued
 valuations $v$.\\This sequent feature, which is only an alternative formulation
 for the 2-valued classic logic, is \emph{fundamental} in the framework of
 many-valued logics where  the semantics for the entailment
 based on the algebraic
 matrices $(X,D)$ is often subjective and  arbitrary. Let us consider, for example, the fuzzy logic with the uniquely \emph{fixed semantics} for all logical connectives,
 where   the subset of designated
 elements $D \subseteq X$ is an arbitrary-subjective choice between the \emph{infinite} number of closed intervals $[a,
 1]$ also for very restricted interval, for example, $0.83 \leq a \leq 0.831$. It does not happen in the classic 2-valued logics where
 a different logic semantics (entailment) is obtained by adopting only different semantics
 for some of its logical connectives, usually for negation operator.
 This property of the classic 2-valued logic can be propagated to
 many-valued logics by adopting the principle of classic
 truth-invariance for the entailment: in that case for fixed semantics
 of all logical connectives of a given language we obtain a unique logic.
 \\
 The definition of the 2-valued entailment in the sequent system $\G$, given in Definition \ref{def:entailment},
 can replace the current entailment based on algebraic
 matrices $(X,D)$ where $D \subseteq X$ is a subset of designated
 elements.
Thus, we are able now to introduce the many-valued valuation-based
(i.e., model-theoretic) semantics for many-valued logics:
\begin{definition}  \cite{Majk08dC} \label{def:manyvaluation}
 A \verb"many-valued"  model-theoretic semantics of a given many-valued logic $\L$, with a Gentzen system $\G = \langle \mathbb{L}, \Vdash
 \rangle$,
 is the semantic deducibility relation
 $~~\models_m~$,
 defined for any $\Gamma = \{s_i = (\Phi_i \vdash \Psi_i)~|~i \in I\}$ and a sequent $~s = (\Phi \vdash \Psi) \in \mathbb{L} \subseteq \L_M \times \L_M~~$
 by : \\
 $~\Gamma \models_m s~~~~$ iff $~~~~$ "all many-valued models of $\Gamma$ are the models of $s$", that is,\\
 iff $~~~~\forall v \in \mathbb{V}_m(~ \forall(\Phi_i \vdash \Psi_i) \in \Gamma(\mathfrak{F}(v)(\Phi_i) \leq \mathfrak{F}(v)(\Psi_i)~$
 implies $~\mathfrak{F}(v)(\Phi) \leq \mathfrak{F}(v)(\Psi))$.
\end{definition}
\begin{lemma} For any $\Gamma = \{s_i = (\Phi_i \vdash \Psi_i)~|~i \in I\}$ and a sequent
 $~s = (\Phi \vdash \Psi) ~~$
 we have that $~\Gamma \models_m s~~~~$ iff $~~~~\forall v \in
 Mod_{\Gamma} (~\mathfrak{B}(v)(s) = 1)$.
\end{lemma}
\textbf{Proof:} We have that $~\Gamma \models_m s~~~~$ iff\\
$~~~~\forall v \in \mathbb{V}_m(~ \forall(\Phi_i \vdash \Psi_i) \in
\Gamma(\mathfrak{F}(v)(\Phi_i) \leq \mathfrak{F}(v)(\Psi_i)~$
 implies $~\mathfrak{F}(v)(\Phi) \leq \mathfrak{F}(v)(\Psi))$\\
 iff $~~\forall v \in
 Mod_{\Gamma} (~ \forall (\Phi_i \vdash \Psi_i) \in \Gamma(\mathfrak{F}(v)(\Phi_i) \leq \mathfrak{F}(v)(\Psi_i)~$ implies $~\mathfrak{F}(v)(\Phi) \leq \mathfrak{F}(v)(\Psi))$\\
 iff $~~~~\forall v \in
 Mod_{\Gamma} (~\mathfrak{F}(v)(\Phi) \leq \mathfrak{F}(v)(\Psi))$\\
iff $~~~~\forall v \in
 Mod_{\Gamma} (~\mathfrak{B}(v)(s) = 1)$.\\$\square$\\
It is easy to verify that any  many-valued logic has a Gentzen-like
system $\G = \langle \mathbb{L}, \Vdash
 \rangle$ (see the definition at the begining of this Section) that is a \emph{normal} logic.
\begin{theo}   \label{prop:manyvaluation}
Many-valued model theoretic semantics is an adequate semantics for a
many-valued logic $\L$ specified by a Gentzen-like logic system $\G
= \langle \mathbb{L}, \Vdash
 \rangle$, that is, it is sound and complete. Consequently,
$~~~\Gamma \models_m s~~$ iff $~~\Gamma \Vdash s$.
\end{theo}
\textbf{Proof:} Let us prove that for any many valued model $v \in
Mod_{\Gamma}$, the obtained sequent bivaluation $~\beta =
\mathfrak{B}(v):\L_M \times \L_M \rightarrow \textbf{2}$ is the
characteristic function of the closed theory $\Gamma_{v} = C(T)~$
with $~~T =  \{ \Phi \vdash 1, ~1 \vdash \Phi~|~\Phi \in P_{mm}, ~
\mathfrak{F}(v)(\Phi) = 1 \}~ \bigcup ~  \{ \Phi \vdash 0, ~0
\vdash \Phi~|~\Phi \in P_{mm}, ~ \mathfrak{F}(v)(\Phi) = 0 \}$.\\
1. Let us show that for any sequent $s$, $~~ s\in \Gamma_v$ implies $~\beta(s) = 1$:\\
First of all, for any sequent $s \in T$ we have that:
  if it is of the form $\Phi \vdash 1$ or $1
\vdash \Phi$ (where $~\Phi \in P_{mm}$) we have that
$~\mathfrak{F}(v)(\Phi) = 1$, thus $s$ is satisfied by $v$ (it holds
that $1 \leq 1$ in both cases); if it is of the form $\Phi \vdash 0$
or $0 \vdash \Phi$ we have that $~\mathfrak{F}(v)(\Phi) = 0$, thus
$s$ is satisfied by $v$ (it holds that $0 \leq 0$ in both cases).
Consequently, \emph{all} sequents in $T$ are satisfied by $v$.\\
From  Proposition \ref{prop:sound} we have that all inference rules
in $\G$ are sound w.r.t. the model satisfiability. Thus for any
deduction $T \Vdash s$ (i.e., $s \in \Gamma_v$) where all sequents
in premises are satisfied by the many-valued valuation (model) $v$,
the deduced sequent $s = (\Phi \vdash \Psi)$ must be satisfied as
well, that is, it must hold that $\mathfrak{F}(v)(\Phi) \leq
\mathfrak{F}(v)(\Psi)$, i.e., $\beta(s) = 1$.\\
2. Let us show that for any sequent $s$,  $~\beta(s) = 1$ implies
$~~ s\in \Gamma_v$:
 For \emph{any} sequent $s =(\Phi \vdash \Psi) \in \L_M \times \L_M$ if $~\beta(s) =
 1$ we have one of the two possible cases:\\
 2.1 Case when $\mathfrak{F}(v)(\Phi) = 0$. Then (from Proposition
 \ref{prop:norform}) $~\Phi$  can be substituted by $~ \bigvee_{1 \leq i \leq
 m}(\bigwedge_{1 \leq j \leq m_i}
 ([y_{ij1}]...[y_{ijk_{ij}}])A_{ij})$, i.e., $~\mathfrak{F}(v)( \bigvee_{1 \leq i \leq
 m}(\bigwedge_{1 \leq j \leq m_i}
 ([y_{ij1}]...[y_{ijk_{ij}}])\\A_{ij})) = \bigvee_{1 \leq i \leq
 m}\mathfrak{F}(v)(\bigwedge_{1 \leq j \leq m_i}
 ([y_{ij1}]...[y_{ijk_{ij}}])A_{ij}) =
 0$,\\ that is, for every $~1 \leq i \leq m$, $~\mathfrak{F}(v)(\bigwedge_{1 \leq j \leq m_i}
 ([y_{ij1}]...[y_{ijk_{ij}}])A_{ij}) =  0$, i.e.\\ $~(\bigwedge_{1 \leq j \leq m_i}
 ([y_{ij1}]...[y_{ijk_{ij}}]A_{ij}) \vdash 0) \in T$. Consequently, by
 applying the inference rule 2b, we deduce $~T \Vdash (\bigvee_{1 \leq i \leq
 m}(\bigwedge_{1 \leq j \leq m_i}
 ([y_{ij1}]...[y_{ijk_{ij}}])A_{ij})\vdash 0)$, that is, by the substitution inference
 rule (for $~\sigma:  \bigvee_{1 \leq i \leq
 m}(\bigwedge_{1 \leq j \leq m_i}
 ([y_{ij1}]...[y_{ijk_{ij}}])A_{ij}) \mapsto \Phi$) we obtain that $~T \Vdash (\Phi \vdash 0)$.
 From the fact that $~0 \vdash \Psi$ is an axiom in $\G$, and by
 applying the transitive inference rule, we obtain that $~ T \Vdash (\Phi
 \vdash \Psi)$, i.e., $~~s  \in  C(T) = \Gamma_v$.\\
2.2 Case when $\mathfrak{F}(v)(\Phi) = 1$. In this case, from the
fact that this sequent is satisfied, $\mathfrak{F}(v)(\Psi) = 1$
must be true as well. Thus, we can substitute $\Psi$ by $1 \vee
\Psi$, so that we obtain the axiom $1 \vdash 1 \vee \Psi $ in $\G$,
and, consequently,  by applying the substitution inference rule (for
$~\sigma: 1 \vee \Psi \mapsto \Psi$) we obtain $~T \Vdash (1 \vdash
\Psi)$. From the fact that $~\Phi \vdash 1$ is an axiom in $\G$ and
by  applying the transitive inference rule we obtain that $~ T
\Vdash (\Phi
 \vdash \Psi)$, i.e., $~~s  \in  C(T) = \Gamma_v$.\\
So, from (1) and (2) we obtain that $~~~\beta(s) = 1~~$ iff $~~ s\in
\Gamma_v$, i.e., the sequent bivaluation $\beta$
 is the characteristic function of a closed set.
  Consequently, any many-valued \emph{model} $v$ of this many-valued logic $\L$
corresponds to the \emph{closed} bivaluation $\beta$ which is a
characteristic function of a closed theory of sequents:  we define
the set of all closed bivaluations obtained from the set of
many-valued models $v \in Mod_{\Gamma}$: $~~Biv_{\Gamma} =
\{\Gamma_{v}~|~v \in Mod_{\Gamma} \}$. From the fact that $\Gamma$
is satisfied by every $v \in Mod_{\Gamma}$ we have that for every
$\Gamma_v \in Biv_{\Gamma}$, $\Gamma \subseteq \Gamma_v$, so that
$C(\Gamma) = \bigcap Biv_{\Gamma}$ (an intersection of closed sets
is a closed set also).  Thus, for $s = (\Phi \vdash \Psi)$,
$~~~\Gamma \models_m s~~$ iff \\$~~~~\forall v \in
 Mod_{\Gamma} (~ \forall (\Phi_i \vdash \Psi_i) \in \Gamma(\mathfrak{F}(v)(\Phi_i) \leq
 \mathfrak{F}(v)(\Psi_i))~$ implies $~\mathfrak{F}(v)(\Phi) \leq \mathfrak{F}(v)(\Psi))$\\
 iff $~~~~\forall v \in
 Mod_{\Gamma} (~ \forall (\Phi_i \vdash \Psi_i) \in \Gamma(\beta(\Phi_i \vdash
 \Psi_i) = 1)~$ implies $~\beta(\Phi\vdash \Psi) = 1)$\\
 iff $~~~~\forall v \in
 Mod_{\Gamma} (~ \forall (\Phi_i \vdash \Psi_i) \in \Gamma((\Phi_i \vdash
 \Psi_i) \in \Gamma_v)~$ implies $~s \in \Gamma_v)$\\
 iff $~~~~\forall \Gamma_v \in
 Biv_{\Gamma} (~ \Gamma \subseteq \Gamma_v~$ implies $~s \in
 \Gamma_v)$  \\
 iff $~~~~\forall \Gamma_v \in
 Biv_{\Gamma} (~s \in
 \Gamma_v)$ $~~~~$,  because $\Gamma \subseteq \Gamma_v$ for each  $\Gamma_v \in Biv_{\Gamma}$\\
 iff $~~~~~s \in
 \bigcap Biv_{\Gamma} = C(\Gamma)$, that is, iff $~~~~~\Gamma \Vdash s$.
 \\$\square$\\
Thus, in order to define the model-theoretic semantics for
many-valued logics we do not need to use the matrices: we are able
to use only the many-valued valuations, and \emph{many-valued
models} (i.e., the valuations which satisfy all sequents in $\Gamma$
of a given many-valued logic $\L$). This point of view is used also
for the
definition of a new representation theorem for many-valued complete lattice based logics in \cite{Majk06th,Majk06ml}.\\
 Here, in  a many-valued
logic $\L$,  specified by a set of sequents in $\Gamma$, for a
formula $\phi \in \L$ that has the same value $x \in X$ (for
\emph{any} truth value $x$) for all many-valued models $v \in
Mod_{\Gamma}$, we have that its modal version $[x]\phi$ is a
theorem; that is, \\ $~~\forall v\in Mod_{\Gamma} (v(\phi) = x)~~$
iff $~~\Gamma \Vdash (1 \vdash [x]\phi
)$,\\
that corresponds to the \emph{truth-invariance} many-valued
entailment (MV)
 in subsection
\ref{subsec:truth-invariance}. Such a value $x \in X$ does not need
to be a designated element $x \in D$, as in the matrix semantics for
a many-valued logic. This fact explains way we do not need a
semantic specification by matrix designated elements.
\\\\
\textbf{Example 2:} Let us prove that, given an assumption $\Gamma =
\{ 1 \vdash [x]\phi, 1 \vdash [y]\psi \}$, then $[z](\phi \odot
\psi)$ for $z = x\odot y$, is deduced from $\Gamma$, that is $\Gamma
\Vdash ([z](\phi \odot \psi))$; or equivalently if $[x]\phi$ and
$[y]\psi$ are valid (i.e., for every valuation $v \in \mathbb{V}_m$,
$v:\L \rightarrow X$, the values of $\phi$ and $\psi$ are equal to
$x$ and $y$ respectively, i.e. $\forall v \in \mathbb{V}_m (v(\phi)
= x$ and $v(\psi) = y)$), then
$[z](\phi \odot \psi)$ is valid as well.\\
As first step we introduce the equivalence relation $\equiv$, such
that $\Phi \equiv \Psi$ iff $\Phi \vdash \Psi$ and $\Psi \vdash
\Phi$ (i.e., when $\Phi$ iff $\Psi$). Consequently, from the
reflexivity axiom in $\G$ we have that $\Phi \equiv \Phi$, as for
example $1 \equiv 1$. The equivalent formulae can be used in the
substitution inference rule: if $\Phi \equiv \Psi$ then we can use
the substitution of $\Phi$ by $\Psi$, that is, the substitution
$\sigma:\Phi \mapsto \Psi$.\\
Let us show the simple equivalence $\Phi \vee \Phi \equiv \Phi$: \\
from the reflexivity axiom $\Phi \vdash \Phi$, by using the upper
bound inference rule (when $\Psi = \Phi$) we deduce $\Phi \vee \Phi
\vdash \Phi$. Then, from the axioms for join  $\Phi \vdash
\Phi \vee \Phi$, we obtain $\Phi \vee \Phi \equiv \Phi$.\\
Now, from the assumptions $1 \vdash [x]\phi, 1 \vdash [y]\psi \in
\Gamma$, by the lower bound inference rule in $\G$, we obtain (a) $1
\vdash [x]\phi \wedge [y]\psi$, i.e., $\Gamma \Vdash (1 \vdash
[x]\phi \wedge [y]\psi)$. Let $z = x\odot y$ and let us denote
$[x]\phi \wedge [y]\psi)$ by $\Phi$, so that (a) becomes (a')
$1\vdash \Phi$. Now we can take the axiom for join, (b) $\Phi \vdash
\Phi \vee \bigvee_{v,w \in X. v\odot w = z }([v]\phi \wedge
[w]\psi)$, so from (a') and (b) and the transitivity rule we obtain
$1\vdash \Phi \vee \bigvee_{v,w \in X. v\odot w = z} ([v]\phi \wedge
[w]\psi)$, i.e., (c) $1 \vdash \Phi \vee \Phi \vee \Psi$, where
$\Psi = \bigvee_{v,w \in X ( v \neq x, w \neq y, v\odot w = z)}
([v]\phi \wedge [w]\psi)$. Thus, by substitution $\sigma: \{ 1
\mapsto 1, \Phi \vee \Phi \mapsto \Phi \}$ and by applying the
substitution rule to (c) we deduce the sequent $1 \vdash \Phi \vee
\Psi $ , that is (d) $1\vdash
  \bigvee_{v,w \in X. v\odot w = z} ([v]\phi \wedge [w]\psi)$
  (from $\Phi \vee \Psi = \bigvee_{v,w \in X,
v\odot w = z} ([v]\phi \wedge [w]\psi)$). Consequently, by applying
the transitivity rule to the sequent (d) and the introduction axiom
$\bigvee_{v,w \in X, v\odot w = z} ([v]\phi \wedge [w]\psi) \vdash
[z](\phi \odot \psi)$, we deduce $1 \vdash [z](\phi \odot \psi)$,
that is $\Gamma \Vdash (1 \vdash [z](\phi \odot \psi))$.\\
Notice that in such deductions no one of the values $x, y, z$ has to
be designated value. We do not make any distinction for the
 truth values in $X$.
\\$\square$\\
 \textbf{Remark:} There is also  another way,
alternative to 2-valued sequent systems, to reduce the many-valued
logics into "meta" 2-valued logics: it is based on  Ontological
encapsulation \cite{Majk04on,Majk06MV}, where each many-valued
proposition (or many-valued ground atom $p(a_1, ..,a_n)$) is
ontologically encapsulated into the "flattened" 2-valued atom
$p_F(a_1, ..,a_n, x)$ (by enlarging original atoms with a new logic
variable  whose domain of values is the set of truth values of the
complete lattice $x \in X$: roughly, "$p(a_1, ..,a_n)$ has a value
$x$" iff  $p_F(a_1, ..,a_n, x)$ is true).
 In fact, such a flattened atom is logically equivalent to the  sequent $(1 \vdash [x]p(a_1, ..,a_n))$.
\section{Autoreferential Kripke-style semantics} \label{Section:Kripke}
 We are able to
 define an equivalence relation $\approx_L$ between the formulae of any many-valued logic
  based on the set of  truth values $X$, in order to
 define the Lindenbaum algebra for this logic, $(\L/\approx_L) $,
 where for any two formulae $~\phi, \psi \in \L$,  $~~~~\phi \approx_L
 \psi~~$ iff  $~~\forall v \in \mathbb{V}_m (v(\phi) = v(\psi))$.\\
 Thus, the elements of this quotient algebra $\L/\approx_L$ are the
 equivalence classes, denoted by $\langle \phi \rangle$.\\
  In particular we will consider an  equivalence class $\langle \phi
\rangle$ (the set
 of all equivalent formulae to $\phi$ w.r.t. $\approx_L$)  that has exactly one
 constant $x \in X$, which is an element of this equivalence class (we abuse a denotation here by denoting by $x$ a formula,
  such that has a constant logic value $x \in X$ for
 every interpretation $v$,  as well), and we
 can use it as the representation element for this equivalence
 class $\langle x
\rangle$. Thus, every formula in this equivalence class has the same
 truth-value as this constant.
 Consequently, we have the injection $i_X:X \rightarrow \L/_{\approx_L}$ between elements in the complete lattice
 $(X, \leq)$ and elements in the Lindenbaum algebra, such that for
 any logic value  $x \in X$,  we obtain the equivalence class $\langle
 x \rangle = i_X(x)  \in \L/_{\approx_L}$.
 It is easy to extend this injection into a monomorphism between the
 original algebra and this Lindenbaum algebra, by definition of
 correspondent connectives in this Lindenbaum algebra. For example:
 $\langle x \wedge y\rangle = i_X(x \wedge y) = i_X(x) \wedge_L i_X(y) = \langle
 x \rangle \wedge_L \langle y
\rangle$, $\langle \neg x \rangle = i_X(\neg x) = \neg_L i_X(x) =
\neg_L\langle x \rangle$,
 etc..\\
In an autoreferential semantics \cite{Majk06ml,Majk08dC,MaPr09} we
will assume that each equivalence class of formulae $\langle \phi
\rangle$ in this Lindenbaum algebra corresponds to one "state -
description". In particular, we are interested to the subset of
"state - descriptions" that are \emph{invariant} w.r.t. many-valued
interpretations $v$, so that can be used as the possible worlds in
the Kripke-style semantics for the original many-valued modal logic.
But from the injection $i_X$ we can take for such an invariant
"state -description" $\langle x \rangle \in \L/_{\approx_L}$ only
its inverse image  $x = i_X^{-1}(\langle x
\rangle) \in X$.\\
 %
%
 Consequently, the
set of possible worlds in this
autoreferential semantics corresponds to the set of truth values $X$.\\
Now we will consider the Kripke model (introduced in subsection
\ref{subsection1}) for the 2-valued multi-modal logic language
$\L_M$ given in Definition \ref{def:multimodal}, obtained from the
many-valued predicate logic language $\L_P$ (defined at the
beginning of Section \ref{section2}):
\begin{definition}\textsc{Kripke semantics}:  \label{def:transSem}
 Let $\L_P$ be a many-valued predicate logic language, based on a set $X$ of  truth values,  with a set of predicate letters $P$ and  Herbrand base $H$.
 Let  ${\M_v} = (F, S, V)$ be a Kripke model of its
correspondent 2-valued multi-modal logic language $\L_M^*$ with the
frame $F = (\W, \{{\R}_w = \W \times \{ w\} ~|~w \in \W\})$ where
$\W = X \bigcup \textbf{2}$ and with mapping $~~V:\W \times P
\rightarrow {\bigcup}_{n < \omega} \textbf{2}^{S^n}$, such that for
any n-ary predicate $p \in P$, and tuple $(c_1,..,c_n) \in S^n$,
there exists a unique $w \in \W$, such that $V(w,p)(c_1,..,c_n) = 1$.   \\
It defines the Herbrand interpretation $v:H \rightarrow X$, such
that $v(p(c_1,..,c_n)) = w$ iff $V(w,p)(c_1,..,c_n) = 1$, and its
unique homomorphic extension to all ground formulae $v:\L
\rightarrow X$.\\
  Let $~g:Var \rightarrow S~$ be an assignment
 for object variables  $\nu_i \in Var$, $i = 1,2,..$,
and  $w \in \W$,
 then the  satisfaction relation $~\models~$ is defined by $~~\M_v\models_{g,w} \phi~~$ iff
$~~v(\phi/g) = w$, for any many-valued formula $\phi \in \L_P$. It is extended to all modal formulae in $\L_M^*$ as follows:\\
1. $~~\M_v\models_{g,w}1 ~~$ and $~~\M_v\nvDash_{g,w}0 $, for tautology and contradition respectively.\\
2. $~~\M_v\models_{g,w}[x]\Phi~~$ iff $~~\forall y((w,y)\in \R_{x}$
implies $~ \M_v\models_{g,y}\Phi)$, $~~$ for any  $\Phi \in \L_M^*$ or  $\Phi \in \L_P$.\\
3. $~\M_v\models_{g,w} \Phi \wedge \Psi ~~$ iff $~~\M_v\models_{g,w}
\Phi ~$ and $~\M_v\models_{g,w} \Psi $, $~~$ for $\Phi, \Psi \in \L_M^*$.\\
4. $~\M_v\models_{g,w} \Phi \vee \Psi ~~$ iff $~~\M_v\models_{g,w}
\Phi ~$ or $~\M_v\models_{g,w} \Psi $, $~~$ for $\Phi, \Psi \in
\L_M^*$.
\end{definition}
Notice that, based on this Kripke model $\M_v$, it is defined a
many-valued valuation $v:H \rightarrow X$, with a unique standard
homomorphic extension $v:\L \rightarrow X$, as follows from
definition above: for any ground atom $p(c_1,..,c_n)\in H$, we
define $v(p(c_1,..,c_n)) = w$ where $w \in
X$ is a unique value which satisfies $~~V(w, p)(c_1,..,c_n) = 1$.\\
Vice versa, given a many-valued model $v:H\rightarrow X$ for a
many-valued predicate logic language $\L_P$, we define a Kripke
model with mapping $V$ such that for any $w \in \W$, n-ary $p \in
P$, and a tuple $(c_1,..,c_n) \in S^n$, $~~V(w, p)(c_1,..,c_n) = 1~$
iff
$~v(p(c_1,..,c_n)) = w$.\\ $\square$\\
 Let $~\Psi/g \in \L_M$ be a ground formula
obtained from $\Psi \in \L_M^*$ by assignment $g$, then we denote by
$\|\Psi/g\|$ the set of worlds where the ground formula $\Psi/g \in
\L_M$, is satisfied, with
 $~\|p(\nu_1,..,\nu_n)/g\| = \{v(p(g(\nu_1),..,g(\nu_n)))\} $ and $~\|\phi/g\| = \{v(\phi/g))\} $, $\phi \in \L$.\\
Thus, differently from the original many-valued ground atoms in $\L$
which can be satisfied only in one single world,  the modal atoms in
$\L_M$ have the standard 2-valued property, that is, they are true
or false in these Kripke models, and, consequently, are satisfiable
in all possible worlds, or absolutely not satisfiable in any world.
Thus, our positive multi-modal logic with modal atoms $\L_M$
satisfies the classic 2-valued properties:
\begin{propo} \label{Prop:2-valued}For any ground formula $\Phi/g$ of the positive multi-modal
logic $\L_M$ defined in Definition \ref{def:multimodal}, we have
that $\|\Phi/g\| \in \{\emptyset, \W\}$, where $\emptyset$ is the
empty set.
\end{propo}
\textbf{Proof}: By structural induction :\\
1. $~\|1\| = \W~~$ and $~\|0\| = \emptyset$.\\
2. $~\|[x]\phi/g\| = \W~~$ if $~x = v(\phi/g)$; $~~\emptyset$, otherwise.\\
Let $\Phi, \Psi$ be the two atomic modal formulae  such that,  by
inductive hypothesis $\|\Phi/g\|, \|\Psi/g\| \in \{\emptyset,
\W\}$. Then,\\
3. $~\| [x]\Phi \|  = \{ w \in \W~|~~x \in \|\Phi/g\|\} = \W$ if  $\|\Phi/g\| = \W$; $~~\emptyset$ otherwise.\\
4. $~ \|(\Phi \wedge \Psi)/g\| = \|\Phi/g\| \bigcap \|\Psi/g\| \in \{\emptyset, \W \}$.\\
5. $~ \|(\Phi \vee \Psi)/g\| = \|\Phi/g\| \bigcup \|\Psi/g\| \in
\{\emptyset, \W \}$.
\\ Thus, from the fact that \verb"any" formula $~\Phi \in \L_M^*$  is logically equivalent
to  disjunctive modal formula $~~ \bigvee_{1 \leq i \leq
 m}(\bigwedge_{1 \leq j \leq m_i}
 ([y_{ij1}]...[y_{ijk_{ij}}])A_{ij})$ (from Proposition  \ref{prop:norform}),
 where each $A_{ij} \in H$ is a ground atom (such that by inductive hipothesis for any modal atom it holds that
 $\|([y_{ij1}]...[y_{ijk_{ij}}])A_{ij}/g \| \in \{\emptyset, \W \}$), and from points 3 and 4 above, we obtain that\\ $\| \Phi/g \| =
 \| \bigvee_{1 \leq i \leq
 m}(\bigwedge_{1 \leq j \leq m_i}
 ([y_{ij1}]...[y_{ijk_{ij}}])A_{ij}/g)\| \in \{\emptyset, \W \}$.
 \\ $\square$ \\
 The following proposition demonstrates the existence of a
one-to-one correspondence between the unique many-valued model of a
 many-valued logic  $\L$ and the Kripke model of a multi-modal
positive logic  $\L_M$.
\begin{propo} \label{prop:completeness} For any
many-valued formula $\phi/g \in \L_P$, we have that $~~v(\phi/g) = x
~~$ iff $~~\mathfrak{F}(v)([x]\phi/g) = 1 ~~$ iff $~~\|[x]\phi/g\| =
\W$.
\end{propo}
\textbf{Proof:} For any ground formula, we have that
$~~\mathfrak{F}(v)([x]\phi/g) = 1 ~~$ iff (by Proposition
\ref{prop:reduction}) $~~\mathfrak{F}(v)(\widehat{[x]\phi/g}) = 1
~~$ iff (by Proposition \ref{prop:preserving}) $~~v(\phi/g) = x $.\\
Thus, it is enough to prove that $~~\alpha(\widehat{[x]\phi/g}) = 1
~~$ iff $~~\|[x]\phi/g\| = \W$, where $\alpha = \mathfrak{F}(v)$. In
the first step, proceeding from left to right, we will demonstrate
it by structural induction on the length (number of logic
connectives) of the
formula $\phi$ : \\
1. The simplest case when $\phi/g = p(c_1,..,c_k)$, $c_i = g(\nu_i)
\in S$, $1 \leq i \leq k$,  is a ground atom for the k-ary predicate
letter $p \in P$. Then,  if $\alpha(\widehat{[x]\phi/g}) =
\alpha(\widehat{[x]p(c_1,..,c_k)}) = \alpha([x]p(c_1,..,c_k)) = 1$,
then (by Proposition \ref{prop:preserving}) $x = v(p(c_1,..,c_k))$,
and from Definition
\ref{def:transSem} we have that $\|[x]\phi/g\| = \W$.\\
2. Let us now suppose, by inductive hypothesis, that it holds for
all formulae with N logical connectives in $\Sigma$. Then, for any
formula $\phi \in \L_P$ with
$N+1$ logical connectives we have the following two cases:\\
2.1. Case  when $\phi = \sim \phi_1$ where $\sim \in \Sigma$ is a
unary connective. Then, if $\alpha(\widehat{[x]\phi/g}) = $ (from
Definition \ref{def:reduction}) $ = \alpha(\bigvee_{y \in X. x =
\sim y} \widehat{[y]\phi_1/g}) =$ (from Proposition
\ref{prop:reduction}) $ = \alpha(\bigvee_{y \in X. x = \sim y}$ \\
$[y]\phi_1/g) =$ (from the homomorphism of $\alpha$) $ = \bigvee_{y
\in X. x = \sim y} \alpha([y]\phi_1/g) = 1$. Thus, there exists $y
\in X$ such that $~ x = \sim y$ and $\alpha([y]\phi_1/g) = 1$. That
is, from Proposition \ref{prop:reduction},
$\alpha(\widehat{[y]\phi_1/g}) = 1$, and from the inductive
hypothesis for this $y$ we obtain (a) $\|[y]\phi_1/g\| = \W$. So, we
obtain $\|[x]\phi/g\| = \|\bigvee_{y \in X. x = \sim y}
\widehat{[y]\phi_1/g}\| = $ (from the point 5 in Definition
\ref{def:transSem}) $ = \bigcup_{y \in X. x = \sim y}\|[y]\phi_1/g\|
= $ (from (a)) $ = \W$.\\
 2.2. Case when $\phi =  \phi_1 \odot \phi_2$, where $\odot \in
\Sigma$ is a binary connective.\\ Then, if
$\alpha(\widehat{[x](\phi_1 \odot \phi_2)/g}) = $ (from Definition
\ref{def:reduction}) $ = \alpha(\bigvee_{y,z \in X. x =  y \odot z}
(\widehat{[y]\phi_1/g} \wedge \widehat{[z]\phi_2/g})) \\ =$ (from
Proposition \ref{prop:reduction}) $ = \alpha(\bigvee_{y,z \in X. x =
y \odot z} ([y]\phi_1/g \wedge [z]\phi_2/g)) =$ (from the
homomorphism of $\alpha$) $ = \bigvee_{y,z \in X. x =  y \odot z}
(\alpha([y]\phi_1/g) \wedge \alpha([z]\phi_2/g)) = 1$. Then, there
exist $y,z \in X$ such that $~ x =  y \odot z$ with
$\alpha([y]\phi_1/g) =1 $ and $\alpha([z]\phi_2/g) = 1$. That is,
from Proposition \ref{prop:reduction} $\alpha(\widehat{[y]\phi_1/g})
= 1$, $~\alpha(\widehat{[z]\phi_2/g}) = 1$, and from inductive
hypothesis for this $y$ we obtain $\|[y]\phi_1/g\| = \W$ and
$\|[z]\phi_2/g\| = \W$, that is (b) $\|[y]\phi_1/g \wedge
[z]\phi_2/g\| = \|[y]\phi_1/g\| \bigcap \|[z]\phi_2/g\| = \W$. So,
we obtain $\|[x]\phi/g\| = \|\bigvee_{y,z \in X. x =  y \odot z}
(\widehat{[y]\phi_1/g} \wedge \widehat{[z]\phi_2/g})\| = $ (from the
point 5 in Definition \ref{def:transSem}) $ = \bigcup_{y,z \in X. x
=  y \odot z}
\|[y]\phi_1/g \wedge [z]\phi_2/g\| =$ (from (a)) $ = \W$.\\
Consequently, we have shown that $~~\alpha(\widehat{[x]\phi/g}) = 1
~~$ implies $~~\|[x]\phi/g\| = \W$. Vice versa, the proof from right
to left is analogous. \\Let us show now that for any ground formula
$\Phi/g \in \L_M$ it holds
that $\alpha(\Phi/g) = 1 ~~$ iff $~~\|\Phi/g\| = \W$:\\
From the fact that  \verb"any" $~\Phi \in \L_M$  is logically
equivalent to  disjunctive modal formula $~\bigvee_{1 \leq i \leq
 m}(\bigwedge_{1 \leq j \leq m_i}
 ([y_{ij1}]...[y_{ijk_{ij}}])A_{ij})$, we obtain that if,\\ $\alpha(\Phi/g) =
\alpha(~\bigvee_{1 \leq i \leq
 m}(\bigwedge_{1 \leq j \leq m_i}
 ([y_{ij1}]...[y_{ijk_{ij}}])A_{ij}/g)) = \\ =~\bigvee_{1 \leq i \leq
 m}(\bigwedge_{1 \leq j \leq m_i}
 \alpha(([y_{ij1}]...[y_{ijk_{ij}}])A_{ij}/g)) = 1$, \\ then there exists $i$, ($1 \leq i
\leq m$), such that for all $1 \leq j \leq m_i$,
$\alpha(([y_{ij1}]...[y_{ijk_{ij}}])A_{ij}/g) = 1$, i.e.,
$\|([y_{ij1}]...[y_{ijk_{ij}}])A_{ij}/g\| = \W$.  Thus,
$\|\bigwedge_{1 \leq j \leq m_i}
 ([y_{ij1}]...[y_{ijk_{ij}}])A_{ij}/g \| = \bigcap_{1 \leq j \leq m_i}
 \|([y_{ij1}]...[y_{ijk_{ij}}])A_{ij}/g)\ \| = \W$, and consequently,
$~~\|\Phi/g\| =  \W$. \\And vice versa.
\\$\square$
\begin{coro} Given a many-valued model $v$ for a many-valued logic language $\L_P$, for any ground formula $\Phi/g \in \L_M$ holds that $~~\mathfrak{F}(v)(\Phi/g)
= 1 ~~$ iff $~~\|\Phi/g\| = \W$, (i.e.,  $~\Phi/g$ is true in the
Kripke model in Definition \ref{def:transSem}).
\end{coro}
\textbf{Proof:} By structural recursion and by Propositions
\ref{Prop:2-valued} and \ref{prop:completeness}.
\\$\square$\\
 From this corollary  we
obtain  that any true formula $\Phi \in \L_M$ is true also in the
Kripke model, and vice versa. That is, the autoreferential
Kripke-stile semantics for the multi-modal logic $\L_M$, defined in
Definition \ref{def:transSem}, is \emph{sound} and \emph{complete}.
\section{Conclusion}
The  main goal of this paper has been the development of a new
\emph{binary} sequent  calculi, with truth-invariance entailment,
for a many-valued predicate logic language $\L_P$ with a finite set
of truth values $X$, and the definition of  Kripke-like semantics
for it, both sound and complete. We did not use any ordering of
truth values in $X$ nor any algebraic matrix with a strict subset of
designated truth values. So, from this point of view, it is the most
general semantic approach for the many-valued logics. In more
specific cases, when the set $X$ is a complete lattice of truth
values, there is also another non-matix based approach (with
truth-preserving entailment) presented in \cite{Majk06ml,Majk06th}
where the sequent system is based on the lattice poset of truth values.\\
In comparison with the standard historical approach based on
m-sequents, this approach is deterministic, in the way that the
axiomatic sequent system is uniquely determined by the set of
many-valued logical connectives. It is  more compact and is a
particular implementation of standard two-sided  sequent systems,
where the left side of each sequent is just a single formula, as
well. Moreover, it is not matrix-based and does need any definition
of a subset of designated elements in $X$, as in other approaches
where for each subjectively defined subset of designated truth
values (consider for example the logic with $n = 10^{30}$
truth-values, $X = \{\frac{i }{n}~|~ 1\leq i \leq n\} $), for the
same logic language $\L$ and the same semantics for its logical
connectives, we obtain the different deductive system. Here this
subjectiveness   is avoided, based on the generalization of the
2-valued truth-invariance principle for the logic entailment, and
the resulting deductive
system for a many-valued logic with fixed semantics of its logical connectives is general and  uniquely defined as in all cases of the 2-valued logics.
\\
Differently from other approaches we defined also a Kripke-like
semantics for this many-valued deductive system, because our
encapsulation of many-valued logic into the 2-valued sequent system
is based on the introduction of the finite set of  modal operators,
for each  truth value in $X$, that is, this 2-valued encapsulation
is modal as in \cite{Majk06MV}. The frame in this autoreferential
Kripke semantics, based on the Lindenbaum algebra considerations, is
finite and uniquely determined by the set of truth values in $X$.


\bibliographystyle{IEEEbib}
\bibliography{mydb}

\end{document}